\documentclass[aps,prc,superscriptaddress,showpacs,secnumroman,showkeys]{revtex4}
\usepackage{amsmath,bm}%
\usepackage{mathrsfs}
\usepackage{graphics}
\usepackage{graphicx}
\usepackage{color}%
\usepackage{float}
\usepackage{color}
\begin{document}

\def\Journal#1#2#3#4{{#1} {{#2}}, {#3} (#4).}
\def\ANP{Adv. Nucl. Phys.}
\def\ARNPS{Ann. Rev. Nucl. Part. Sci.}
\def\CTP{Commun. Theor. Phys.}
\def\EPJA{Eur. Phys. J. A}
\def\EPJC{Eur. Phys. J. C}
\def\IJMPA{International Journal of Modern Physics A}
\def\IJMPE{International Journal of Modern Physics E}
\def\JCHP{J. Chem. Phys.}
\def\JCP{Journal of Computational Physics}
\def\JHEP{JHEP}
\def\JPCS{Journal of Physics: Conference Series}
\def\JPG{J. Phys. G: Nucl. Part. Phys.}
\def\NATURE{Nature}
\def\NC{La Rivista del Nuovo Cimento}
\def\NCA{IL Nuovo Cimento A}
\def\NPA{Nucl. Phys. A}
\def\NST{Nuclear Science and Techniques}
\def\PA{Physica A}
\def\PAN{Physics of Atomic Nuclei}
\def\PHY{Physics}
\def\PRA{Phys. Rev. A}
\def\PRC{Phys. Rev. C}
\def\PRD{Phys. Rev. D}
\def\PLA{Phys. Lett. A}
\def\PLB{Phys. Lett. B}
\def\PLD{Phys. Lett. D}
\def\PRL{Phys. Rev. Lett.}
\def\PL{Phys. Lett.}
\def\PREV{Phys. Rev.}
\def\PREP{\em Physics Reports}
\def\PROG{Progress in Particle and Nuclear Physics}
\def\RPP{Rep. Prog. Phys.}
\def\RDNC{Rivista del Nuovo Cimento}
\def\RMP{Rev. Mod. Phys}
\def\SCIENCE{Science}
\def\ZPA{Z. Phys. A.}

\def\ANN{Ann. Rev. Nucl. Part. Sci.}
\def\ANNAST{Ann. Rev. Astron. Astrophys.}
\def\AP{Ann. Phys}
\def\APJ{Astrophysical Journal}
\def\APJS{Astrophys. J. Suppl. Ser.}
\def\EJP{Eur. J. Phys.}
\def\LANC{Lettere Al Nuovo Cimento}
\def\NCA{Nuovo Cimento A}
\def\PHYS{Physica}
\def\NP{Nucl. Phys}
\def\MATH{J. Math. Phys.}
\def\JPAM{J. Phys. A: Math. Gen.}
\def\PRO{Prog. Theor. Phys.}
\def\NPB{Nucl. Phys. B}

\title{Description of Charged Particle Pseudorapidity Distributions in Pb+Pb Collisions with Tsallis Thermodynamics}
\author{Y. Gao}
\affiliation{School of Information Engineering, Hangzhou Dianzi University, Hangzhou 310018, People's Republic of China;}
\author{H. Zheng}
\email[]{zheng@lns.infn.it}
\affiliation{Laboratori Nazionali del Sud, INFN, I-95123 Catania, Italy;}
\author{L.L. Zhu}
\affiliation{College of Physical Science and Technology, Sichuan University, Chengdu 610064, People's Republic of China;}
\author{A. Bonasera}
\affiliation{Laboratori Nazionali del Sud, INFN, I-95123 Catania, Italy;}
\affiliation{Cyclotron Institute, Texas A$\&$M University, College Station, Texas 77843, USA.}
\date{\today}

\begin{abstract}
The centrality dependence of pseudorapidity distributions for charged particles produced in Au+Au collisions at  $\sqrt{s_{NN}}=130$ GeV and 200 GeV at RHIC, and in Pb+Pb collisions at $\sqrt{s_{NN}}=2.76$ TeV at LHC are investigated in the fireball model, assuming that the rapidity axis is populated with fireballs following one distribution function. We assume that the particles in the fireball fulfill the Tsallis distribution. The theoretical results are compared with the experimental measurements and a good agreement is found. Using these results, the pseudorapidity distributions of charged particles produced in Pb+Pb central collisions at $\sqrt{s_{NN}}=5.02$ TeV and 10 TeV are predicted. 
\end{abstract}



\maketitle

\section{Introduction}
Relativistic heavy-ion collisions have been used to study the systems with hadronic or partonic degrees of freedom under extremely high temperature and density. The experiments carried out at the Relativistic Heavy-Ion Collider (RHIC) and the Large Hadron Collider (LHC) have attracted much experimental and theoretical interest to study particle production. The pseudorapidity distribution of charged particles is one of the quantities which can be measured directly in experiments. It has the same importance as other quantities, such as the particle spectra. Though the charged particle pseudorapidity distribution can not provide immediate understanding of the particle production mechanism, it is indispensable, as a benchmark tool, to constrain the models and help us to understand the fundamental processes.

In experiments, the pseudorapidity distributions of charged particles ($\frac{dN_{ch}}{d\eta}$) have been extensively measured for different reaction systems at different collision energies and centralities \cite{alice13_2760, alice16_2760, alice16_5020, foka17, alver11}. Several parameterizations have been adopted to describe $\frac{dN_{ch}}{d\eta}$ and extrapolate the total number of charged particles ($N_{ch}$) produced in the reactions. The recently measured results from Pb+Pb at $\sqrt{s_{NN}}=2.76$ TeV have been compared with the ones from the well established  models, such as HIJING \cite{hijingwang}, AMPT \cite{amptlin} (with and without string melting), EPOS-LHC \cite{eposlhc}, UrQMD \cite{urqmd}, CGC based model \cite{cgc}, having been introduced in the field of high energy heavy-ion collisions. None of them can successfully describe the measured distributions except the CGC based model which only gives the results around the mid-pseudorapidity region  \cite{alice13_2760, foka17}. This means more work needs to be done in these sophisticated models even though they have achieved a great success to describe other measurable quantities, such as anisotropic flow.

In Refs. \cite{liu13, liu14}, a multi-source thermal model has been applied to describe the charged particle pseudorapidity distribution on centrality in Pb+Pb at $\sqrt{s_{NN}}=2.76$ TeV. In this model, the contributions to the $\frac{dN_{ch}}{d\eta}$ come from the projectile and target cylinders and the leading particles. The results show that the contributions of leading particles are large and necessary. In Ref. \cite{liu15}, Liu and Gao proposed a new revised Landau hydrodynamic model following the same philosophy as in the multi-source thermal model to systematically study $\frac{dN_{ch}}{d\eta}$ produced in heavy-ion collisions. The system is consisted by the central, target, and projectile three sources. The central source is described by the Landau hydrodynamic model and the other two are generated with the Monte Carlo approach. The contribution ratio from target and projectile sources is smaller than the one from the leading particles in the multi-source thermal model \cite{liu13, liu14, liu15}. Jiang {\it et al.} adopted the 1+1 dimensional hydrodynamics, which is analytically solvable,  to study the pseudorapidity distributions in different collision systems at currently available energies \cite{jiang16}. In this model, the final particles have been classified into two classes: the particles governed by the hydrodynamics and the leading particles. These models have different scenarios, but their results are in good agreement with the experimental data.

Recently and for the first time, the non-extensive approach was used to describe the charged particle rapidity distributions produced in high energy proton-proton (p+p) collisions \cite{cleymans15}. The encouraging results stimulate us to apply the same method to the $\frac{dN_{ch}}{d\eta}$ produced in heavy-ion collisions. This work will be complementary to Ref. \cite{cleymans15} and a new application to the non-extensive distribution whose origin is still under investigation in high energy physics  \cite{cleymans15, hua17, hua15, hua16, huaprd16, cleymans12epja, marques13prd, cleymans16raaepja, cleymans16csepja, wong2012, wong15prd, wilk00prl, star2007, phenix2011, alice1, aliceS2012, cms3, maciej, sena}.

The paper is organized as follows. In section II, we introduce the framework of the fireball model and formulas used in the non-extensive approach. In section III, the prediction of the pseudorapidity distribution of charged particles produced in p+p collisions at 13 TeV has been checked. We also show the results of pseudorapidity distributions of charged particles produced in Pb+Pb at $\sqrt{s_{NN}}=2.76$ TeV and the total number of charged particles at each centrality studied. The pseudorapidity distributions of charged particles produced at different centralities in Au+Au collisions at  $\sqrt{s_{NN}}=130$ GeV and 200 GeV at RHIC are also investigated. Using the results obtained from Au+Au collisions and Pb+Pb collisions at $\sqrt{s_{NN}}=2.76$ TeV, we make the predictions for the pseudorapidity distributions of charged particles produced in Pb+Pb central collisions at $\sqrt{s_{NN}}=5.02$ TeV and 10 TeV. A brief conclusion is given in section IV.

\section{Formula of pseudorapidity distribution}
Recently, the Tsallis distribution has attracted much experimental and theoretical interest because of its great success to describe the particle spectra produced in pp, pA and AA collisions \cite{cleymans15, hua17, hua15, hua16, huaprd16, cleymans12epja, marques13prd, cleymans16raaepja, cleymans16csepja, wong2012, wong15prd, wilk00prl, star2007, phenix2011, alice1, aliceS2012, cms3, maciej, sena}. One needs to notice that in the literature there are different versions of the Tsallis distribution. We will use the non-extensive approach of Ref. \cite{cleymans15} to conduct our study in which the particle distribution can be written as:
\begin{eqnarray}
E\frac{d^3N}{dp^3} = gV\frac{m_T \cosh y}{(2\pi)^3} [1+(q-1)\frac{m_T\cosh y-\mu}{T}]^{-\frac{q}{q-1}}, \label{tsallisB}
\end{eqnarray}
where $g$ is the degeneracy of the particle state, $V$ is the volume, $m_T=\sqrt{m_0^2+p_T^2}$ is the transverse mass and $m_0$ is the particle mass, $y$ is the rapidity, $\mu$ is the chemical potential, $T$ is the temperature and $q$
is the entropic factor which measures the non-additivity of the entropy. The self-consistency of the thermodynamical description has been taken into account \cite{cleymans12epja}. In Eq. (\ref{tsallisB}), there are four parameters, namely $V, \mu, T, q$. $\mu$ will be assumed to be 0 in the following since we will focus on the Pb+Pb collisions at $\sqrt{s_{NN}}=2.76$ TeV and Au+Au collisions at  $\sqrt{s_{NN}}=130$ GeV and 200 GeV. In the mid-rapidity $y=0$ region, Eq. (\ref{tsallisB}) is reduced to
\begin{equation}
E\frac{d^3N}{dp^3} = gV\frac{m_T}{(2\pi)^3} [1+(q-1)\frac{m_T}{T}]^{-q/(q-1)}, \label{tsallisBR}
\end{equation}
which will be used to fit the particle spetrum to extract the paramters $q$ and $T$ in the next section.

The framework used in Ref. \cite{cleymans15} is to assume that the rapidity axis is populated with fireballs following a distribution function given by $\nu (y_f)$, where $y_f$ is the rapidity of the fireball. Particles will appear when the fireballs freeze out and follow the Tsallis distribution Eq. (\ref{tsallisB}). Therefore the particle distribution in terms of transverse momentum and rapidity is given by 
\begin{equation}
\frac{d^2N}{p_Tdp_Tdy} = \frac{N}{A} \int_{-\infty}^{\infty} \nu(y_f)\frac{m_T\cosh(y-y_f)}{(2\pi)^2}[1+(q-1)\frac{m_T\cosh(y-y_f)}{T}]^{-\frac{q}{q-1}}dy_f, \label{eq1}
\end{equation}
where $N$ is the total multiplicity of the particles and $A$ is a normalization constant to ensure $\int_{-\infty}^\infty\int_0^\infty \frac{d^2N}{dp_Tdy}=N$. To obtain the particle distribution only in term of rapidity, one should integrate Eq. (\ref{eq1}) over the transverse momentum $p_T$ and obtain
\begin{eqnarray}
\frac{dN}{dy}&=&\frac{N}{A} \int_{-\infty}^{\infty} dy_f \nu(y_f) T [1+m_0 (q-1)\frac{\cosh(y-y_f)}{T}]^{-\frac{1}{q-1}} \nonumber\\
&&\times \frac{-(q-2)m_0^2+2m_0T \textrm{sech}(y-y_f)+2T^2 \textrm{sech}^2(y-y_f)}{4\pi^2(q-2)(2q-3)}.\label{eqndndy}
\end{eqnarray}
The details can be found in the appendix. It is easy to show that our derivation is the same as the equation (6) in Ref. \cite{cleymans15} but in a simpler form.

Since the experimental data are usually measured in the pseudorapidity ($\eta$) space, we need to convert the formula of the charged particle rapidity distribution to $\frac{dN}{d\eta}$. 
Applying the relation
\begin{equation}
\frac{dy}{d\eta}(\eta, p_T)=\sqrt{1-\frac{m_0^2}{m_T^2 \cosh^2 y}},
\end{equation}
we can rewrite Eq. (\ref{eq1}) as
\begin{equation}
\frac{d^2N}{p_Tdp_Td\eta \sqrt{1-\frac{m_0^2}{m_T^2 \cosh^2 y}}} = \frac{N}{A} \int_{-\infty}^{\infty} \nu(y_f)\frac{m_T\cosh(y-y_f)}{(2\pi)^2}[1+(q-1)\frac{m_T\cosh(y-y_f)}{T}]^{-\frac{q}{q-1}}dy_f. \label{eqn1}
\end{equation}
Unlike the formula of the particle rapidity distribution, we can not integrate the above equation analytically over $p_T$. Therefore the formula of the pseudorapidity distribution is
\begin{eqnarray}
\frac{dN}{d\eta} &=& \frac{N}{A} \int_0^\infty  p_T\sqrt{1-\frac{m_0^2}{m_T^2 \cosh^2 y}} dp_T \int_{-\infty}^{\infty} \nu(y_f)\frac{m_T\cosh(y-y_f)}{(2\pi)^2}[1+(q-1)\frac{m_T\cosh(y-y_f)}{T}]^{-\frac{q}{q-1}}dy_f \nonumber\\
&=& \frac{N}{A}\int_{-\infty}^{\infty}dy_f \int_0^\infty dp_T p_T\sqrt{1-\frac{m_0^2}{m_T^2 \cosh^2 y}} \nu(y_f)\frac{m_T\cosh(y-y_f)}{(2\pi)^2}[1+(q-1)\frac{m_T\cosh(y-y_f)}{T}]^{-\frac{q}{q-1}}, \nonumber\\
\label{eqn2}
\end{eqnarray}
with the relation
\begin{equation}
y = \frac{1}{2}\ln \Big[\frac{\sqrt{p_T^2\cosh^2 \eta+m_0^2}+p_T \sinh \eta}{\sqrt{p_T^2\cosh^2 \eta+m_0^2}-p_T \sinh \eta}\Big].
\end{equation}

Before we can study the $\frac{dN}{d\eta}$, the fireball distribution function $\nu(y_f)$ should be known. In Ref. \cite{cleymans15}, it has been shown that the q-Gaussian function and Gaussian function give the similar results. Therefore, we only adopt the q-Gaussian function, 
\begin{eqnarray}
\nu(y_f) &=& G_{q'}(y_0, \sigma; y_f) + G_{q'}(-y_0, \sigma; y_f) \nonumber\\
&=& \frac{1}{\sqrt{2\pi}\sigma} [1+(q'-1)\frac{(y_f-y_0)^2}{2\sigma^2}]^{-\frac{1}{q'-1}} +  \frac{1}{\sqrt{2\pi}\sigma} [1+(q'-1)\frac{(y_f+y_0)^2}{2\sigma^2}]^{-\frac{1}{q'-1}}. 
\end{eqnarray}
In principle, $q'$ can be different from $q$ for the particle spectrum in Eq. (\ref{tsallisB}), but $q'=q$ is assumed in the following analysis. $y_0$ and $\sigma$ are fitting parameters which will be determined by the experimental data.

\begin{figure} [H]  
        \centering
        \begin{tabular}{c}
        \includegraphics[scale=0.3]{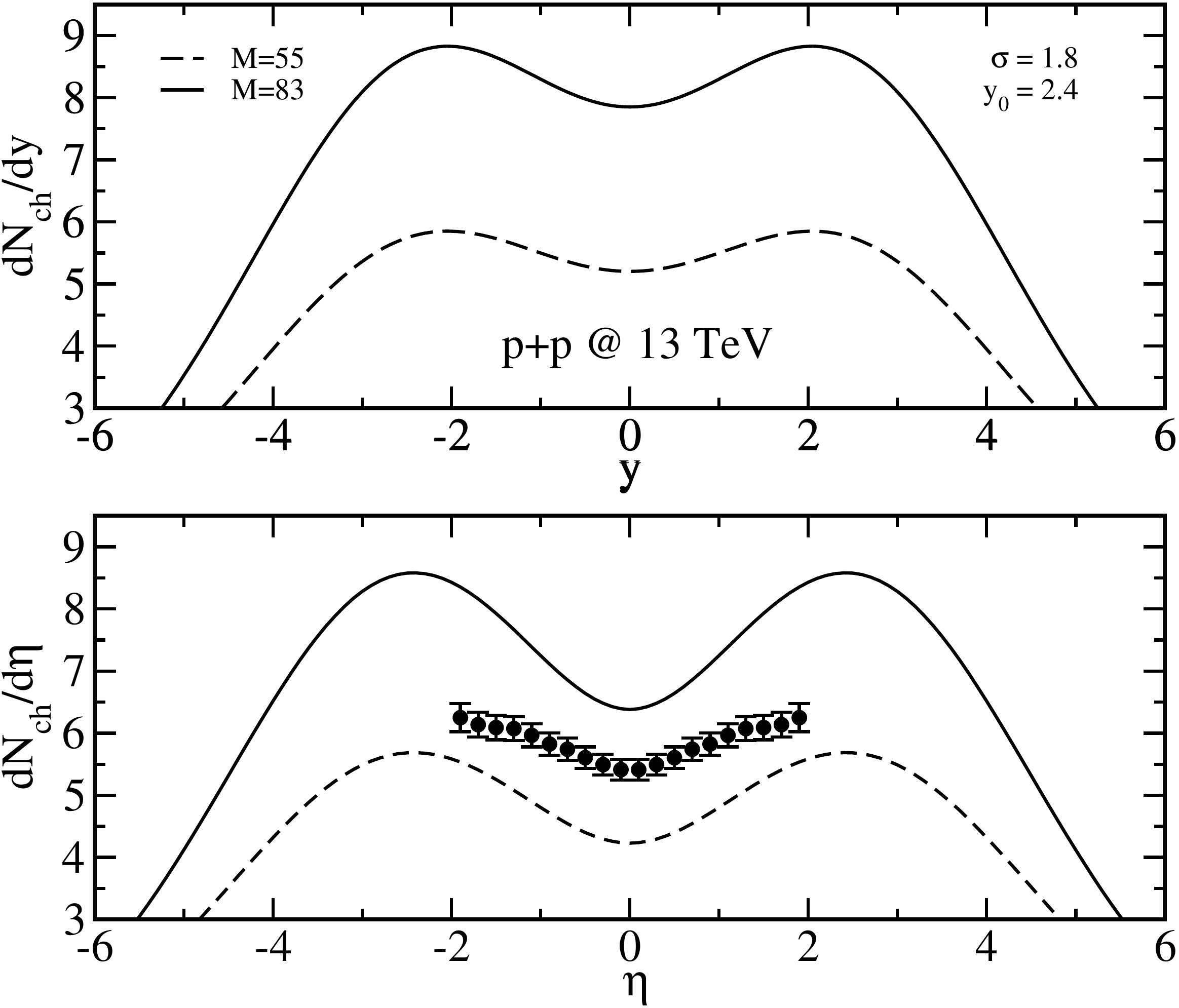}
\end{tabular}
\caption{(Top) the prediction of the charged particle rapidity distribution from Ref. \cite{cleymans15}. (Bottom) the pseudorapidity distribution of the charged particle using the parameters in Ref.  \cite{cleymans15} with Eq. (\ref{eqn2}).}\label{figure1}
    \end{figure}

\section{Results}
 Before we apply the model to Pb+Pb collisions, it is interesting to check the prediction made in Ref. \cite{cleymans15} for p+p collisions at $\sqrt{s_{NN}}=13$ TeV. The values of the parameters in Ref. \cite{cleymans15} are adopted to be consistent. In Fig. \ref{figure1}(a), the predictions for the charged particle rapidity distribution have been reproduced with total multiplicity $M=55$ and $M=83$ using Eq. (\ref{eqndndy}). Since the experimental data were measured in $\frac{dN_{ch}}{d\eta}$, we have to use Eq. (\ref{eqn2}) to recalculate the results and compare with experimental data. In Fig. \ref{figure1}(b), we show that the experimental data are right between the two predicted limits. This indicates that the model works properly.

\begin{figure} [H]  
        \centering
        \begin{tabular}{c}
        \includegraphics[scale=0.3]{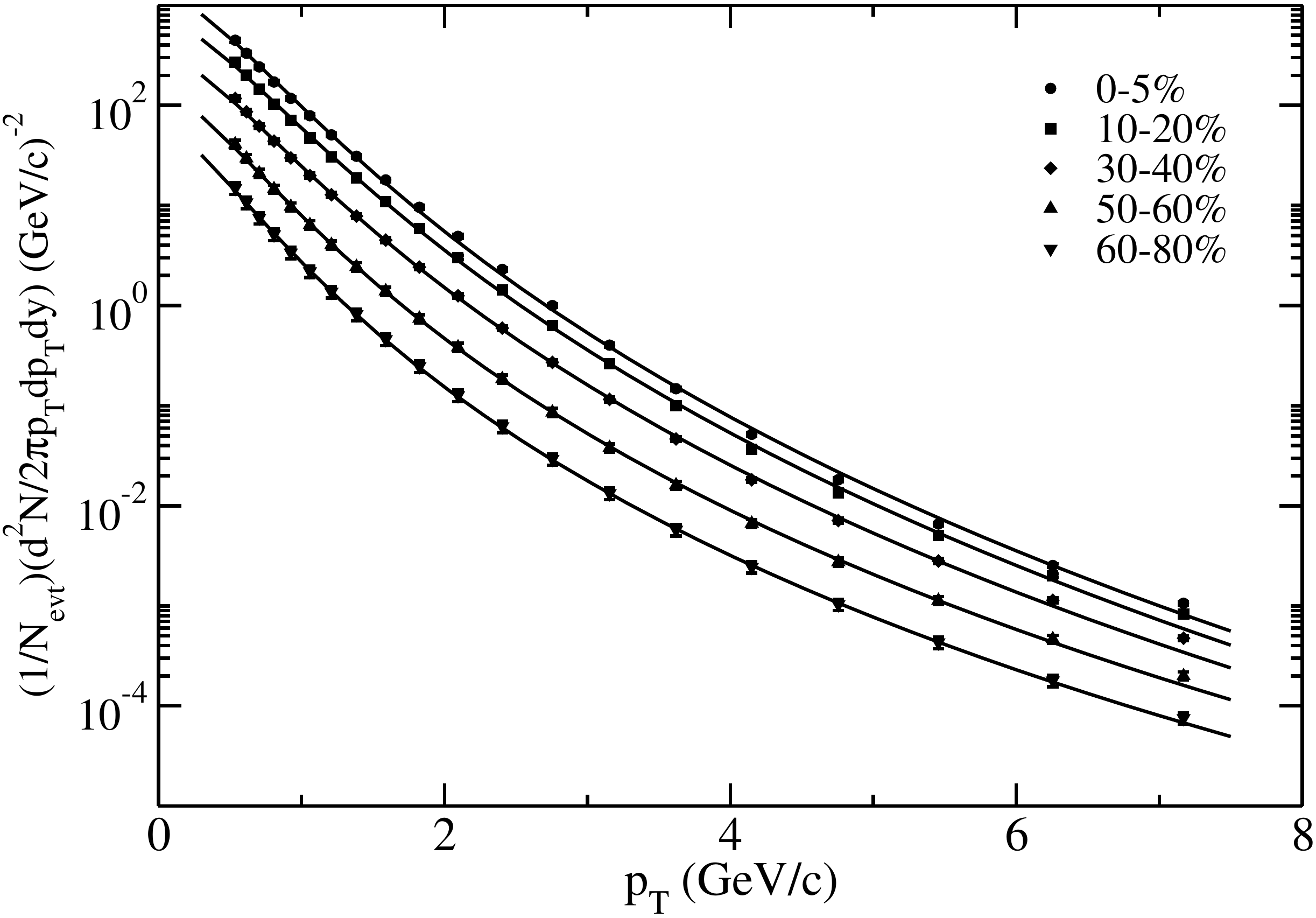}
\end{tabular}
\caption{The lines are the fitting results of the charged particle spetra produced in Pb+Pb at $\sqrt{s_{NN}}=2.76$ TeV with Eq. (\ref{tsallisBR}). The symbols are the experimental data taken from Ref. \cite{atlas15pbpb2760}.}\label{figure2}
    \end{figure}

In Pb+Pb collisions, the situation is different from p+p collisions since many particles are created and the medium plays an important role. As it is shown in Refs. \cite{hua17, hua15, Bylinkin:2014aba,Bylinkin:2015msa, Rybczynski:2014ura}, single Tsallis distribution can not fit the particle spectra produced in Pb+Pb collisions, unlike the p+p collisions.  But the single Tsallis distribution is still able to fit the particle spectra at low and intermediate $p_T$ region where most of the particles are located. It is good enough for us to apply a $p_T$ cut in this study for the charged particle pseudorapidity distribution because most of the particles have been taken into account. In this work, we have chosen the $p_T$ cut at 7.5 GeV/c. In Fig. \ref{figure2} the fitting results of the particle spectra for different centralities in Pb+Pb collisions at $\sqrt{s_{NN}}=2.76$ TeV using Eq. (\ref{tsallisBR}) are shown. As one can see the single Tsallis distribution can reproduce well the experimental data after the $p_T$ cut has been applied to the particle spectrum. The corresponding parameters $T$ and $q$ are extracted. 

\begin{figure} [H]  
        \centering
        \begin{tabular}{c}
        \includegraphics[scale=0.3]{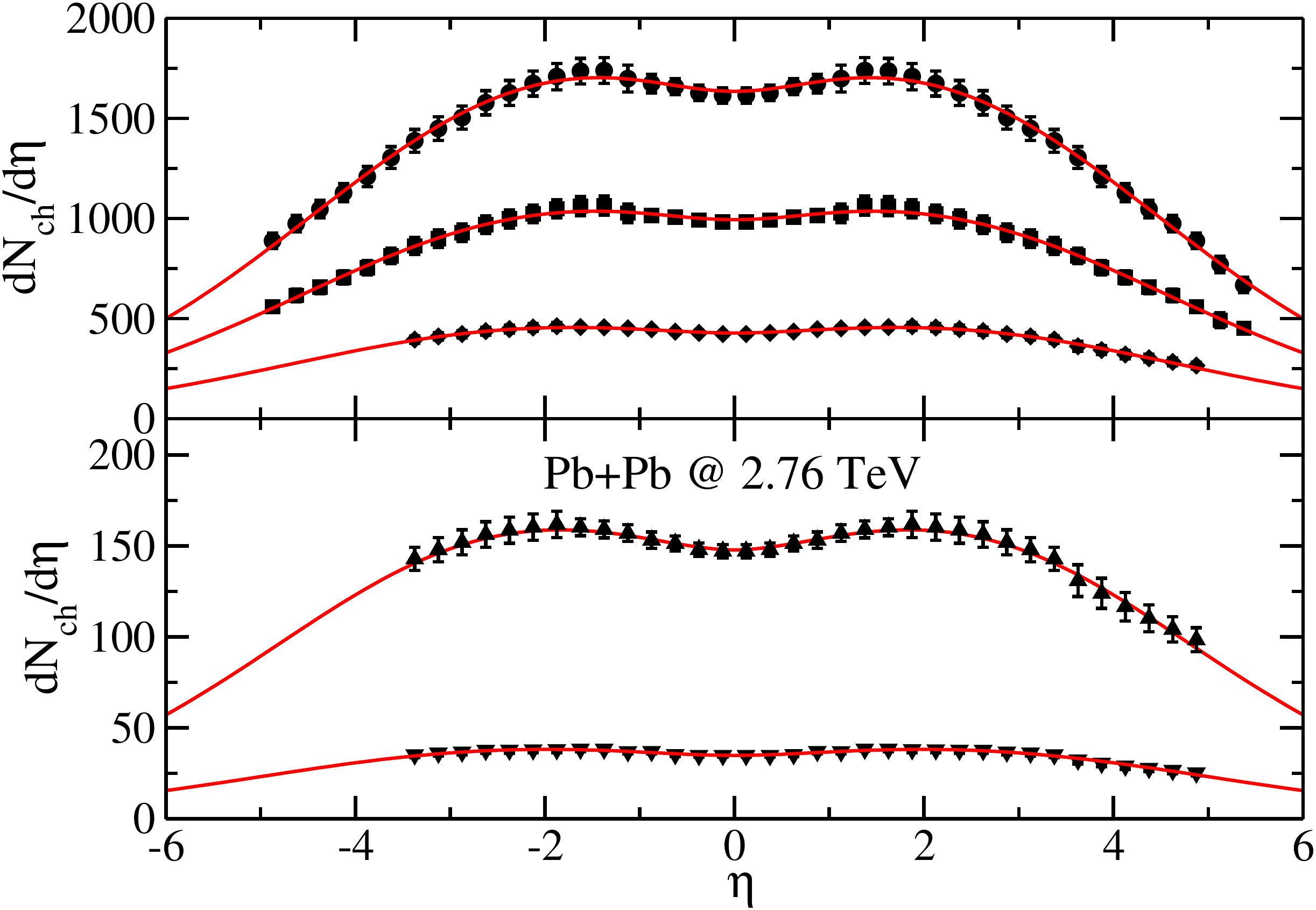}
               \end{tabular}
\caption{(Color online) the charged particle pseudorapidity distribution at centrality (top) 0-5\%, 10-20\%, 30-40\% and (bottom) 50-60\%, 70-80\% produced in Pb+Pb at $\sqrt{s_{NN}}=2.76$ TeV, respectively. The symbols (same as in Fig. \ref{figure2}) are experimental data taken from Ref. \cite{alice16_2760} and curves are the results from Eq. (\ref{eqn2}).}  \label{figure3}
    \end{figure}

After we obtain the values of $T$ and $q$ by fitting the particle spectra at different centralities,  we are able to investigate the charged particle pseudorapdity distribution using Eq. (\ref{eqn2}). In Fig. \ref{figure3}, we plot the results in two panels with different scales in order to better display them. The experimental results are well reproduced in a wide pseudorapidity range with the non-extensive approach proposed in Ref. \cite{cleymans15}.

\begin{figure} [H]  
        \centering
        \begin{tabular}{c}
        \includegraphics[scale=0.3]{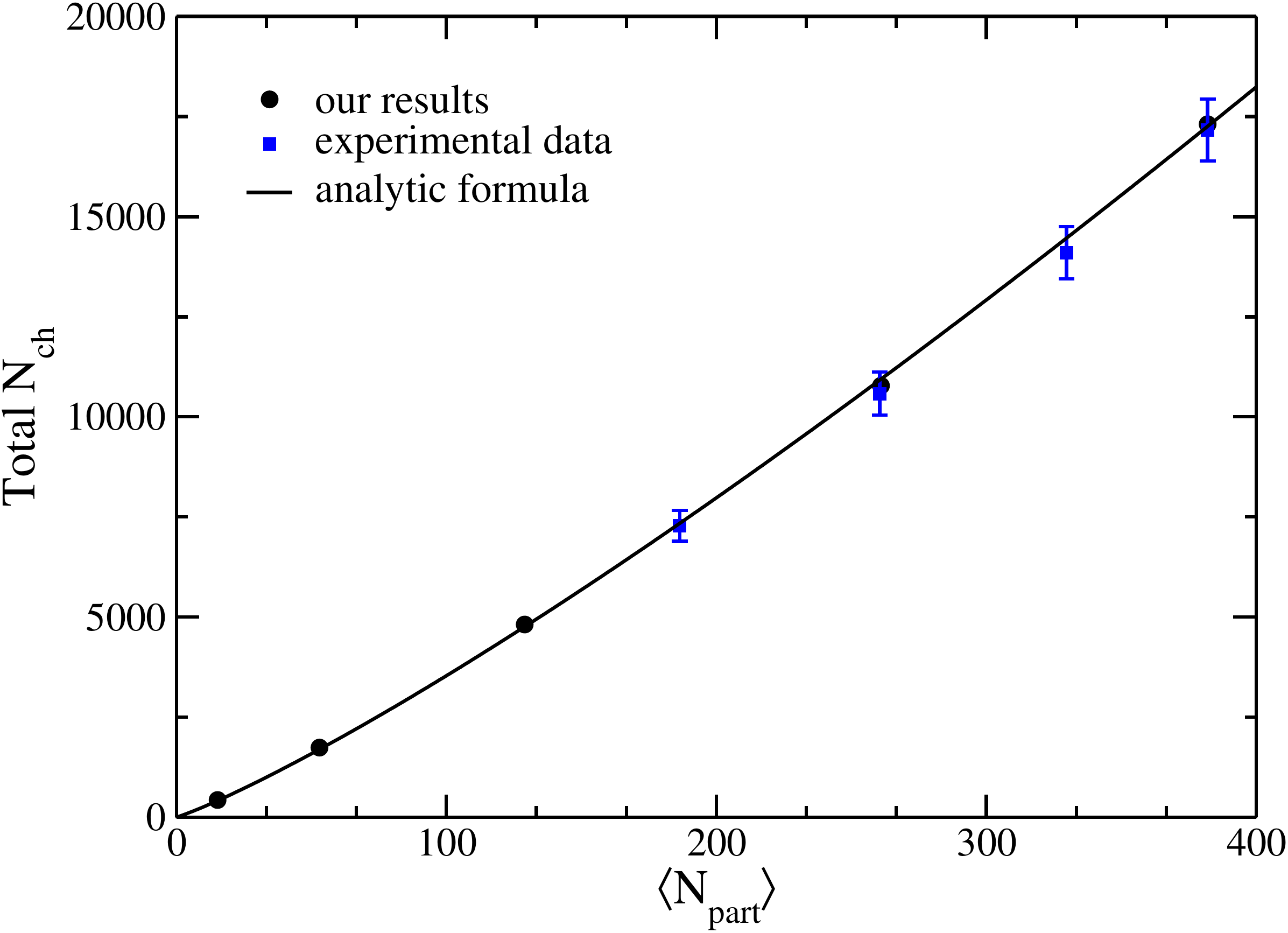}
               \end{tabular}
\caption{(Color online) the total number of charged particles produced in Pb+Pb at $\sqrt{s_{NN}}=2.76$ TeV versus the number of participants. The experimental data are taken from Ref. \cite{alice13_2760} and the analytic curve is taken from Ref. \cite{alice16_2760}.}  \label{figure4}
    \end{figure}
We also are able to estimate the total number of charged particles, integrating the Eq. (\ref{eqn2}) over the pseudorapidity, produced in the Pb+Pb collisions at $\sqrt{s_{NN}}=2.76$ TeV.  In Fig. \ref{figure4}, we show the total number of charged particle obtained from this model with the experimetal data. 

\begin{figure} [H]  
        \centering
        \begin{tabular}{c}
        \includegraphics[scale=0.3]{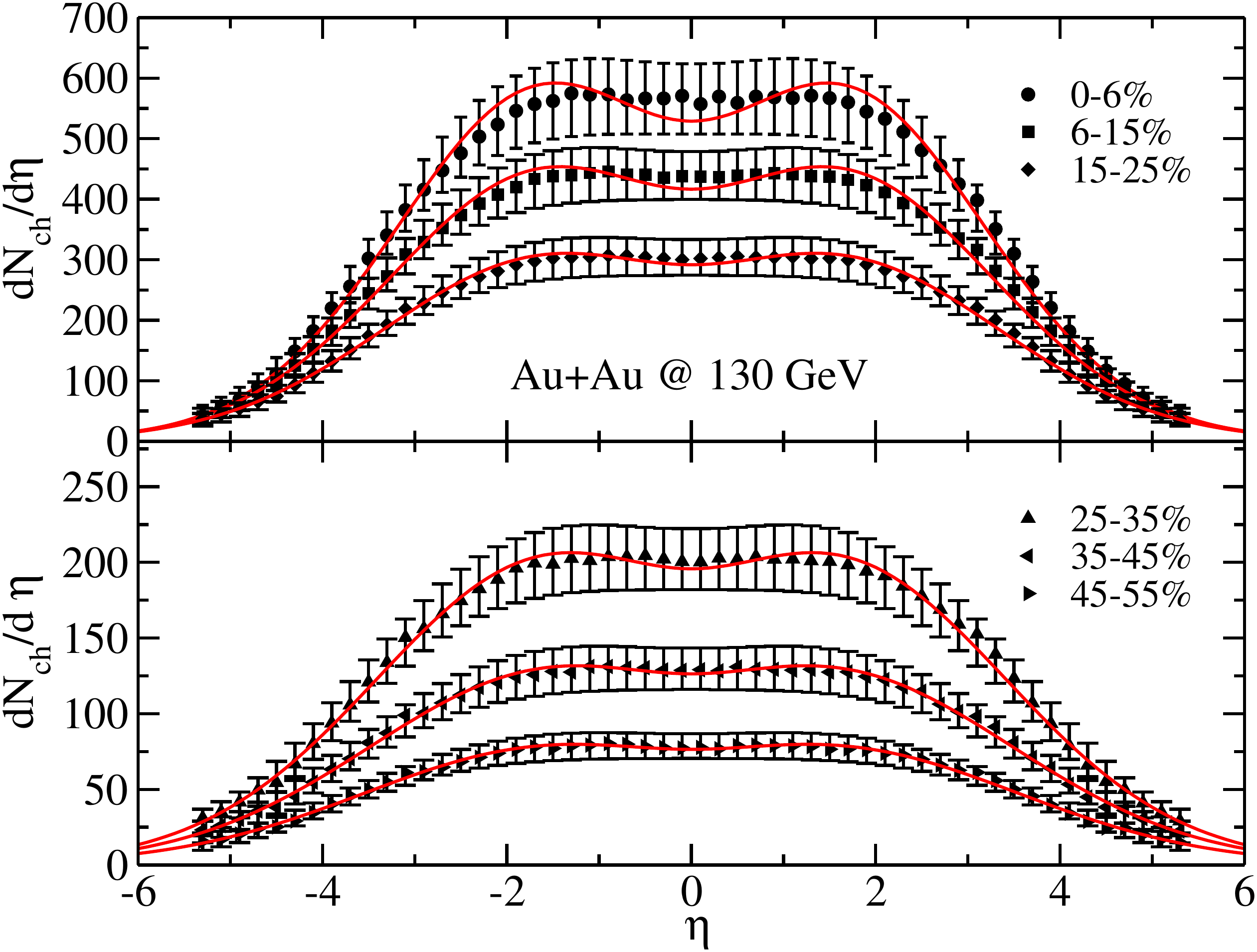}
               \end{tabular}
\caption{(Color online) the charged particle pseudorapidity distribution at centrality (top) 0-6\%, 6-15\%, 15-25\% and (bottom) 25-35\%, 35-45\% and 45-55\% produced in Au+Au at $\sqrt{s_{NN}}=130$ GeV, respectively. The symbols are experimental data taken from Ref. \cite{alver11} and curves are the results from Eq. (\ref{eqn2}).}  \label{auau130}
    \end{figure}

\begin{figure} [H]  
        \centering
        \begin{tabular}{c}
        \includegraphics[scale=0.3]{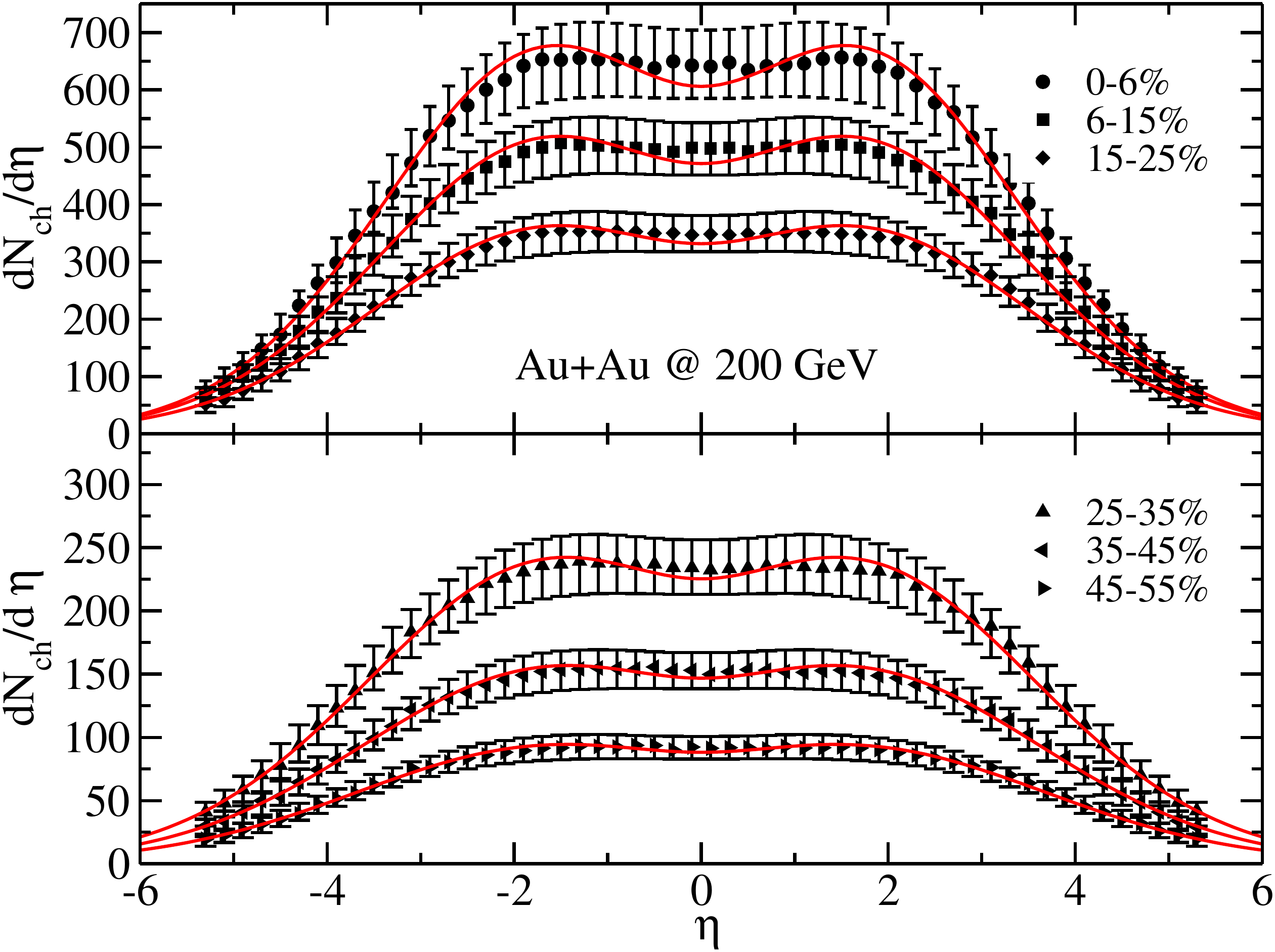}
               \end{tabular}
\caption{(Color online) the charged particle pseudorapidity distribution at centrality (top) 0-6\%, 6-15\%, 15-25\% and (bottom) 25-35\%, 35-45\% and 45-55\% produced in Au+Au at $\sqrt{s_{NN}}=200$ GeV, respectively. The symbols are experimental data taken from Ref. \cite{alver11} and curves are the results from Eq. (\ref{eqn2}).}  \label{auau200}
    \end{figure}

To explore this fireball model, we also apply it to Au+Au collisions at  $\sqrt{s_{NN}}=130$ GeV and 200 GeV respectively. One can repeat the procedure in Pb+Pb collisions at $\sqrt{s_{NN}}=2.76$ TeV and extract the relevant paramters for the model. The particle spectra data are obtained from Refs. \cite{auau130prl, auau200prl}. The results of $\frac{dN_{ch}}{d\eta}$ are shown in Figs. \ref{auau130}, \ref{auau200}. As one can see that the fireball model can reproduce well the pseudorapidity distributions of charged particles produced in Au+Au collsions at different centralities as well.

\begin{figure} [H]  
        \centering
        \begin{tabular}{c}
        \includegraphics[scale=0.3]{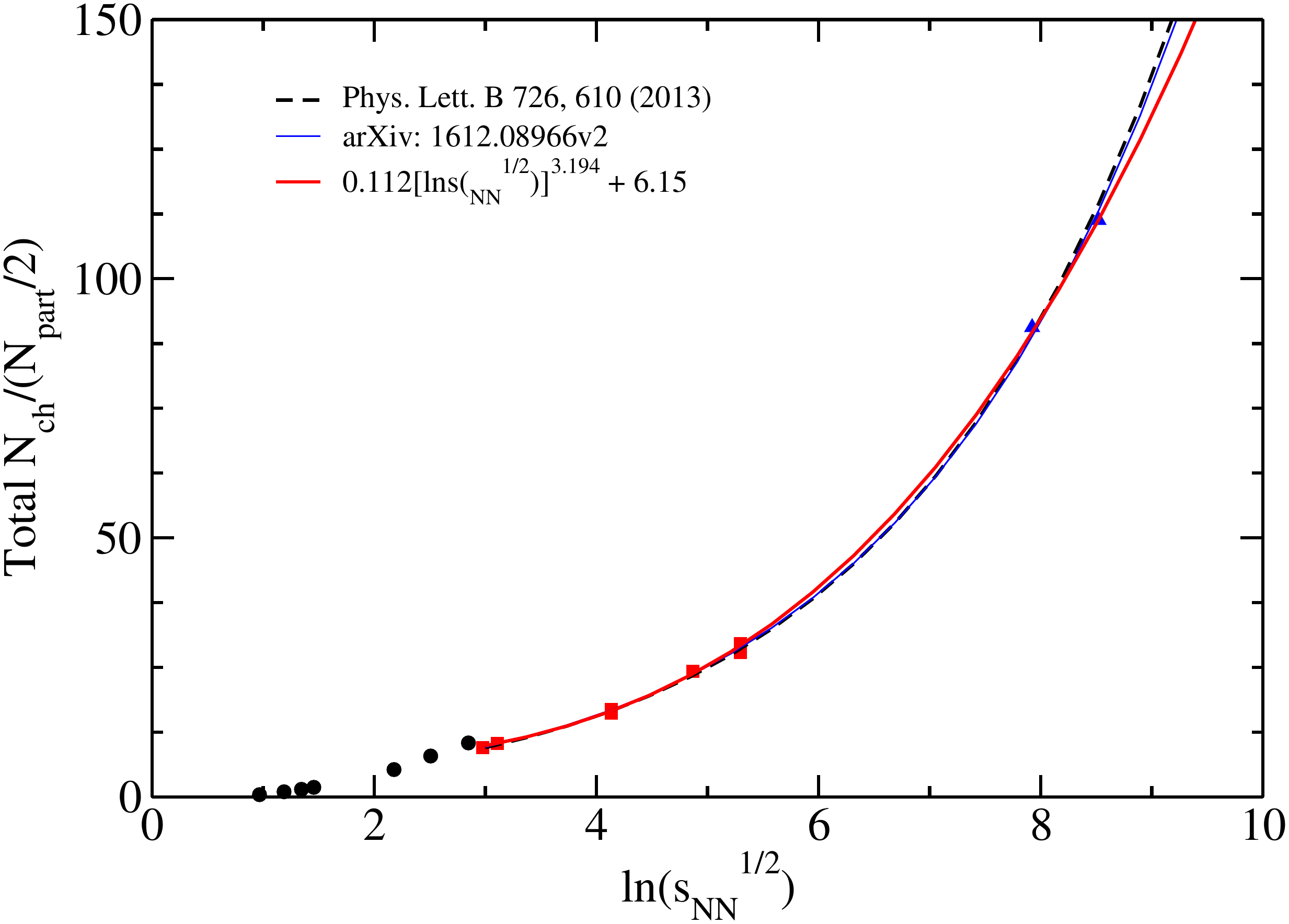}
               \end{tabular}
\caption{(Color online) the total number of charged particles per participant pair produced in the most central collisions versus $\ln(\sqrt{s_{NN}})$. The experimental data at AGS (0-5\% Au+Au), SPS (0-5\% Pb+Pb), RHIC (0-6\% Au+Au and 0-6\% Cu+Cu) and LHC (Pb+Pb 0-5\%) are taken from Refs. \cite{agssps, alver11,alice16_2760, alice16_5020}. The curves are our fitting results and the ones from Refs. \cite{alice13_2760, alice16_5020} to the RHIC and LHC data. }  \label{Nchtot}
    \end{figure}

\begin{figure} [H]  
        \centering
        \begin{tabular}{c}
        \includegraphics[scale=0.3]{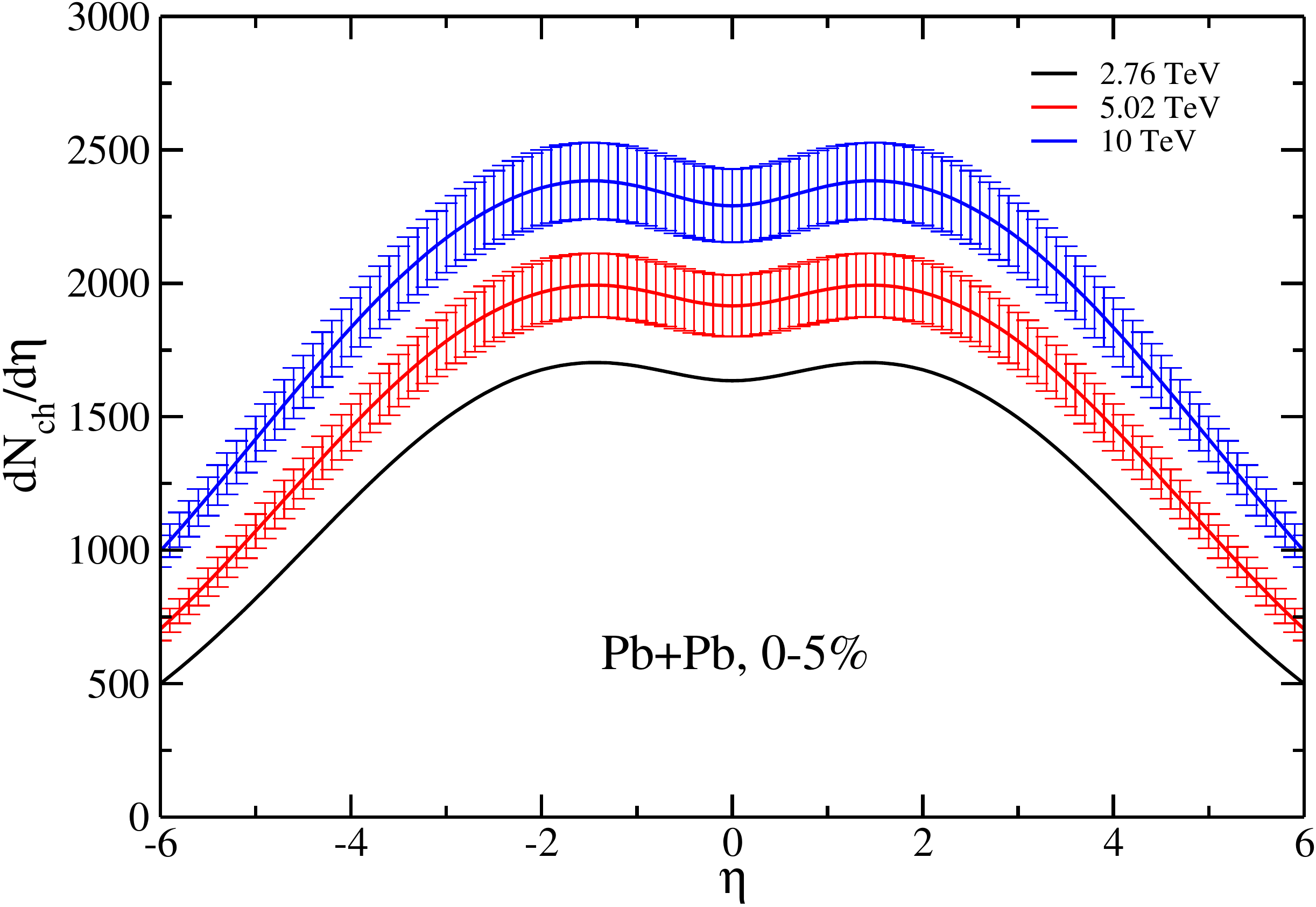}
               \end{tabular}
\caption{(Color online) the predictions of $\frac{dN_{ch}}{d\eta}$ for Pb+Pb central collisions at $\sqrt{s_{NN}}=5.02$ TeV and 10 TeV. The curve for Pb+Pb central collisions at $\sqrt{s_{NN}}=2.76$ TeV is for reference purpose. The error bars for the prediction results are introduced by the uncertanties of the total number of charged particles.}  \label{pbpbsys}
    \end{figure}

Investigating the fitting parameters of $y_0$ and $\sigma$ obtained from Au+Au and Pb+Pb central collisions, we find that they follow a linear relation with $\ln(\sqrt{s_{NN}})$. We can extrapolate to obtain the parameters $y_0$ and $\sigma$ for Pb+Pb collisions at higher energies. This gives us the possibility to predict the charged particle pseudorapidity distributions produced in Pb+Pb collisions. To conduct it, we also need to know the parameters $T$ and $q$ extracted from the particle spectra which are not measured or the data are not available in our prediction energies. We will assume that $T$ and $q$ are the same as the ones obtained in Pb+Pb at $\sqrt{s_{NN}}=2.76$ TeV in our predictions for Pb+Pb at higher energies. Now one more condition is needed which is the total number of charged particles $N_{ch}$. Fortunately, enough experimental data have been accumulated and we can fit them and extrapolate to obtain the $N_{ch}$ at the energy we are interested in. In Fig. \ref{Nchtot}, the experimental results and the fitting results are shown. With these results, we are able to make the predictions for Pb+Pb at $\sqrt{s_{NN}}=5.02$ TeV and 10 TeV. Actually, the particle spectra and $\frac{dN_{ch}}{d\eta}$ have been measured in Pb+Pb collisions at different centralities at $\sqrt{s_{NN}}=5.02$ TeV, but the data are not published yet \cite{alice16_5020}. Since the uncertanties of $N_{ch}$ are 4.5\% at $\sqrt{s_{NN}}=2.76$ TeV and 6.1\% at $\sqrt{s_{NN}}=5.02$ TeV in Pb+Pb central collisions, we add the similar uncertanties 6\% to the extrapolated $N_{ch}$ from our fitting results at our prediction energies. As one can see from Fig. \ref{pbpbsys} and the figure 3 in Ref. \cite{alice16_5020}, our prediction for Pb+Pb at 5.02 TeV is comparable with the experimental data.

\section{Conclusions}
We studied the charged particle pseudorapdity distribution produced in Au+Au collsions at $\sqrt{s_{NN}}=130$ GeV and 200 GeV and in Pb+Pb collisions at  $\sqrt{s_{NN}}=2.76$ TeV using a fireball model with non-extensive approach proposed in Ref. \cite{cleymans15}. In the later case, the total number of charged particles versus centrality is also investigated. The results show that this model is also able to reproduce the experimental results in heavy-ion collisions. It is another contribution to the application of the Tsallis distribution whose origin is still under investigation in high energy physics. Using the data obtained from Au+Au collsions and Pb+Pb collisions, we make the predictions for the pseudorapidity distribution of charged particles produced in  Pb+Pb central collisions at $\sqrt{s_{NN}}=5.02$ TeV and 10 TeV.


\section{Acknowledgements}
The work is supported by the Natural Science Foundation of China (11475050) and the Zhejiang Province science and technology plan project (2015C33035).

\section{Appendix: Formula of rapidity distribution}
 Integrating Eq. (\ref{eq1}) over the transverse momentum $p_T$ yields
\begin{eqnarray}
\frac{1}{N}\frac{dN}{dy}&=&  \frac{1}{A} \int_{-\infty}^{\infty} dy_f \int_0^\infty \nu(y_f) \frac{m_T\cosh(y-y_f)}{(2\pi)^2}[1+(q-1)\frac{m_T\cosh(y-y_f)}{T}]^{-\frac{q}{q-1}}p_T dp_T\nonumber\\
&=& \frac{1}{A} \int_{-\infty}^{\infty} dy_f \nu(y_f) \frac{\cosh(y-y_f)}{(2\pi)^2} \int_0^\infty m_T [1+(q-1)\frac{m_T\cosh(y-y_f)}{T}]^{-\frac{q}{q-1}}p_T dp_T.
\end{eqnarray}
Let us define the second integral as
\begin{equation}
I=\int_0^\infty m_T [1+(q-1)\frac{m_T\cosh(y-y_f)}{T}]^{-\frac{q}{q-1}}p_T dp_T.
\end{equation}
To simplify the integral, we set 
\begin{equation}
a = (q-1)\frac{\cosh(y-y_f)}{T}, \quad b = -\frac{q}{q-1},
\end{equation}
since they do not depend on $p_T$. Therefore
\begin{eqnarray}
I&=& \int_0^\infty m_T (1+am_T)^{b}p_T dp_T \nonumber\\
&=&\int_{m_0}^\infty m_T^2 (1+am_T)^{b}dm_T\nonumber\\
&=&\frac{1}{a^3}\int_{am_0}^\infty x^2(1+x)^b dx\nonumber\\
&=&\frac{1}{a^3} \left\{\frac{1}{b+1}[x^2(1+x)^{b+1}|_{am_0}^\infty - 2\int_{am_0}^\infty x(1+x)^{b+1} dx ]\right\} \nonumber\\
&=& \frac{1}{a^3}\frac{1}{b+1} \left\{-(am_0)^2(1+am_0)^{b+1}-\frac{2}{b+2}[x(1+x)^{b+2}|_{am_0}^\infty - \int_{am_0}^\infty (1+x)^{b+2}dx]\right\} \nonumber\\
&=&\frac{1}{a^3}\frac{1}{b+1} \left[-(am_0)^2(1+am_0)^{b+1}+\frac{2}{b+2}am_0(1+am_0)^{b+2} + \frac{2}{b+2}\int_{am_0}^\infty (1+x)^{b+2}dx\right] \nonumber\\
&=&\frac{1}{a^3}\frac{1}{b+1} \left[-(am_0)^2(1+am_0)^{b+1}+\frac{2}{b+2}am_0(1+am_0)^{b+2} - \frac{2}{(b+2)(b+3)}(1+am_0)^{b+3}\right] \nonumber\\
&=&\frac{1}{a^3}\frac{1}{b+1} (1+am_0)^{b+1} \left[-(am_0)^2+\frac{2}{b+2}am_0(1+am_0) - \frac{2}{(b+2)(b+3)}(1+am_0)^{2}\right] \nonumber\\
&=&\frac{1}{a^3}\frac{1}{b+1} (1+am_0)^{b+1}  \frac{-(b+1)(b+2)(am_0)^2+2(b+1)(am_0)-2}{(b+2)(b+3)} \nonumber\\
&=&\frac{1}{a^3}(1+am_0)^{b+1}  \frac{-(b+1)(b+2)(am_0)^2+2(b+1)(am_0)-2}{(b+1)(b+2)(b+3)}. \label{eqI1}
\end{eqnarray}
The condition that $b$ is a small negative number has been applied during the above calculation. One can calculate 
\begin{equation}
b+1 = -\frac{q}{q-1} + 1 = -\frac{1}{q-1},
\end{equation}
\begin{equation}
b+2 = -\frac{q}{q-1} + 2 = \frac{q-2}{q-1},
\end{equation}
\begin{equation}
b+3 = -\frac{q}{q-1} + 3 = \frac{2q-3}{q-1}.
\end{equation}
Substituting $a$ and these equations into Eq. (\ref{eqI1}), one can obtain
\begin{eqnarray}
I&=&\frac{1}{a^3}(1+am_0)^{-\frac{1}{q-1}}  \frac{\frac{q-2}{(q-1)^2}(am_0)^2-\frac{2}{q-1}(am_0)-2}{-\frac{(q-2)(2q-3)}{(q-1)^3}} \nonumber\\
&=& \frac{(q-1)^3}{a^3}(1+am_0)^{-\frac{1}{q-1}}  \frac{-\frac{q-2}{(q-1)^2}(am_0)^2+\frac{2}{q-1}(am_0)+2}{(q-2)(2q-3)} \nonumber\\
&=&  \frac{(q-1)^3}{[ (q-1)\frac{\cosh(y-y_f)}{T}]^3}[1+m_0 (q-1)\frac{\cosh(y-y_f)}{T}]^{-\frac{1}{q-1}} \nonumber\\
&&\times \frac{-\frac{q-2}{(q-1)^2}[m_0 (q-1)\frac{\cosh(y-y_f)}{T}]^2+\frac{2}{q-1}[m_0 (q-1)\frac{\cosh(y-y_f)}{T}]+2}{(q-2)(2q-3)} \nonumber\\
&=&  \frac{T^3}{\cosh^3(y-y_f)}[1+m_0 (q-1)\frac{\cosh(y-y_f)}{T}]^{-\frac{1}{q-1}} \frac{-(q-2)m_0^2\frac{\cosh^2(y-y_f)}{T^2}+2m_0 \frac{\cosh(y-y_f)}{T}+2}{(q-2)(2q-3)} \nonumber\\
&=&\frac{T}{\cosh^3(y-y_f)}[1+m_0 (q-1)\frac{\cosh(y-y_f)}{T}]^{-\frac{1}{q-1}} \frac{-(q-2)m_0^2\cosh^2(y-y_f)+2m_0T\cosh(y-y_f)+2T^2}{(q-2)(2q-3)}.
\end{eqnarray}
Therefore
\begin{eqnarray}
\frac{1}{N}\frac{dN}{dy}&=& \frac{1}{A} \int_{-\infty}^{\infty} dy_f \nu(y_f) \frac{\cosh(y-y_f)}{(2\pi)^2} \int_0^\infty m_T [1+(q-1)\frac{m_T\cosh(y-y_f)}{T}]^{-\frac{q}{q-1}}p_T dp_T \nonumber\\
&=& \frac{1}{A} \int_{-\infty}^{\infty} dy_f \nu(y_f) \frac{\cosh(y-y_f)}{(2\pi)^2} I \nonumber\\
&=& \frac{1}{A} \int_{-\infty}^{\infty} dy_f \nu(y_f) \frac{\cosh(y-y_f)}{(2\pi)^2}\frac{T}{\cosh^3(y-y_f)}[1+m_0 (q-1)\frac{\cosh(y-y_f)}{T}]^{-\frac{1}{q-1}} \nonumber\\
&&\times \frac{-(q-2)m_0^2\cosh^2(y-y_f)+2m_0T\cosh(y-y_f)+2T^2}{(q-2)(2q-3)} \nonumber\\
&=& \frac{1}{A} \int_{-\infty}^{\infty} dy_f \nu(y_f) T [1+m_0 (q-1)\frac{\cosh(y-y_f)}{T}]^{-\frac{1}{q-1}} \nonumber\\
&&\times \frac{-(q-2)m_0^2+2m_0T \textrm{sech}(y-y_f)+2T^2 \textrm{sech}^2(y-y_f)}{4\pi^2(q-2)(2q-3)}.
\end{eqnarray}
It is easy to show that our derivation is the same as the equation (6) in Ref. \cite{cleymans15} but in a simpler form.






\end{document}